\documentclass[showkeys]{revtex4}
\usepackage{amsmath}
\usepackage{amssymb}
\usepackage{graphicx}
\usepackage[utf8]{inputenc}
\usepackage{color}

\begin{document}

%\newcolumntype{.}{D{.}{.}{-1}}

\title{Wormhole solutions with a complex ghost scalar field and their instability
}

\author{Vladimir Dzhunushaliev}
\email{v.dzhunushaliev@gmail.com}
\affiliation{
	Institute of Experimental and Theoretical Physics,  Al-Farabi Kazakh National University, Almaty 050040, Kazakhstan;
}
\affiliation{
	Department of Theoretical and Nuclear Physics,  Al-Farabi Kazakh National University, Almaty 050040, Kazakhstan;
}
\affiliation{
	Institute of Physicotechnical Problems and Material Science of the NAS of the Kyrgyz Republic, 265 a, Chui Street, Bishkek 720071,  Kyrgyzstan
}
\affiliation{
	Institut f\"ur Physik, Universit\"at Oldenburg, Postfach 2503
	D-26111 Oldenburg, Germany
}

\author{Vladimir Folomeev}
\email{vfolomeev@mail.ru}
\affiliation{
	Institute of Experimental and Theoretical Physics,  Al-Farabi Kazakh National University, Almaty 050040, Kazakhstan;
}
\affiliation{
	Institute of Physicotechnical Problems and Material Science of the NAS of the Kyrgyz Republic, 265 a, Chui Street, Bishkek 720071,  Kyrgyzstan
}
\affiliation{
	Institut f\"ur Physik, Universit\"at Oldenburg, Postfach 2503
	D-26111 Oldenburg, Germany
}

\author{Burkhard Kleihaus}
\email{b.kleihaus@uni-oldenburg.de}
\affiliation{
	Institut f\"ur Physik, Universit\"at Oldenburg, Postfach 2503
	D-26111 Oldenburg, Germany
}

\author{Jutta Kunz}
\email{jutta.kunz@uni-oldenburg.de}
\affiliation{
	Institut f\"ur Physik, Universit\"at Oldenburg, Postfach 2503
	D-26111 Oldenburg, Germany
}

\begin{abstract}
We study compact configurations with a nontrivial
wormholelike spacetime topology
supported by a complex ghost scalar field with a quartic self-interaction.
For this case, we obtain regular asymptotically flat equilibrium solutions possessing reflection symmetry.
We then show their instability with respect to linear radial perturbations.
\end{abstract}

\maketitle

\section{Introduction}

Despite the fact that at the moment there is only one experimentally observed
elementary scalar particle -- the Higgs boson --
it is commonly believed that other types of fundamental scalar fields
do also exist in nature.
Such scalar fields are widely used on microscopic scales in constructing
models of particle physics, they are often considered in
modeling various types of compact astrophysical objects,
and they are important ingredients in describing
the cosmological evolution of the early and the present Universe.
In particular, real and complex scalar fields are employed
in constructing models of gravitating configurations --
the so-called boson stars
\cite{Jetzer:1991jr,Lee:1991ax,Schunck:2003kk,Liebling:2012fv}.
Depending on the masses and self-interactions of the scalar fields,
the resulting boson stars may be microscopic, they may possess
masses and sizes comparable to those of other compact objects,
like neutron stars and black holes; or they may even be suitable to
model the halos of galaxies.

When one considers renormalizable field theories
of real scalar fields in Minkowski space,
it is known that it is impossible to get static regular solutions,
since in a flat four-dimensional spacetime
Derrick's theorem applies \cite{Derrick:1964ww,Rajaraman:1982is}.
Coupling the scalar fields to gravity does not change this basic finding.
There are no localized static  solutions for self-gravitating real scalar
fields; i.e., there are no scalar {\sl geons} \cite{Wheeler:1955zz}.
(Note that we consider neither self-gravitating Skyrmions here,
since the Skyrme model possesses a term quartic in the currents
\cite{Skyrme:1961vq}, nor static scalarons, which possess a
not strictly positive scalar potential
\cite{Kleihaus:2013tba}.)

However, the possibility of obtaining localized static nonsingular solutions
arises when one involves the so-called ghost scalar fields,
i.e., fields with the opposite sign in front of the kinetic term
of the scalar field Lagrangian density.
The possible existence of ghost scalar fields in nature
is indirectly supported by the observed accelerated expansion
of the present Universe
(see, e.g., Refs.~\cite{Perlmutter:1999jt,Bennett:2003bz,Sullivan:2011kv}).
Namely, to explain the recent observational data \cite{Ade:2015xua},
one should take the so-called exotic matter into consideration,
the effective pressure of which is negative and
the modulus of which is greater than its energy density.

One way to introduce such exotic matter is through the use
of ghost scalar fields \cite{Lobo:2005us,Lobo:2017oab}.
On  relatively small scales comparable to the sizes of stars,
such fields permit obtaining solutions of the Einstein-matter equations
describing configurations with a nontrivial wormholelike topology.
There are configurations supported
by massless ghost scalar fields
\cite{Bronnikov:1973fh,Ellis:1973yv,Ellis:1979bh}
and systems possessing a scalar field potential
\cite{Kodama:1978dw,Kodama:1978zg}.
In addition,
such fields allow one to get compact configurations with a trivial spacetime topology
\cite{Dzhunushaliev:2008bq}.

Here, we demonstrate the possibility of obtaining localized
regular solutions with a nontrivial wormholelike topology
supported by a complex ghost scalar field.
To the best of our knowledge, previously, only real ghost scalar fields
have been employed in modeling wormholelike systems.
The use of the
complex ghost scalar field
allows one to introduce a harmonic time dependence  analogously to the case of a complex canonical
scalar field employed in the construction of boson stars.
We also endow the complex ghost scalar field with a nontrivial
quartic self-interaction potential.

The paper is organized as follows. In Sec.~\ref{equilib_confs_gen_eqs},
the general set of equations is derived for the equilibrium configurations
with a nontrivial spacetime topology supported by a complex ghost scalar field.
We present the numerically obtained solutions
for these configurations in Sec.~\ref{equilib_confs_num_sol}.
Subsequently, a linear stability analysis is performed for these solutions
in Sec.~\ref{stab_analysis}.
Finally, in
Sec.~\ref{concl}, we summarize the obtained results.

\section{Equilibrium configurations}
\label{equilib_confs}

\subsection{General set of equations}
\label{equilib_confs_gen_eqs}

We consider here a model of a gravitating complex ghost scalar field.
We start from the action (hereafter, we work in units where $c=\hbar=1$)
\begin{equation}
\label{action_wh_complex}
S=\int d^4 x\sqrt{-g}\left[
-\frac{1}{16\pi G}R
+\frac{1}{2}\left[\varepsilon g^{\mu\nu}\partial_{\mu}\Phi^*\partial_{\nu}\Phi
-V(|\Phi|^2)\right] \right].
\end{equation}
Here, $\Phi$ is a complex scalar field with the potential $V(|\Phi|^2)$,
and $\varepsilon=+1$ or $-1$ corresponds to canonical or ghost fields,
respectively.
This action is invariant under a global phase transformation
$\Phi\to e^{i \theta} \Phi$, which implies the
conservation of its generator $N$
corresponding to the total particle number.

By varying \eqref{action_wh_complex} with respect to the metric,
one obtains the Einstein equations with
the energy-momentum tensor %can be presented as
 \begin{equation}
\label{emt_wh_complex}
T_\nu^\mu=\frac{\varepsilon}{2}g^{\mu\sigma}\left(
\partial_{\sigma}\Phi^*\partial_{\nu}\Phi+\partial_{\sigma}\Phi\partial_{\nu}\Phi^*
\right)-\frac{1}{2}\delta^\mu_\nu\left(
\varepsilon g^{\lambda \sigma}\partial_{\lambda}\Phi^*\partial_{\sigma}\Phi-V
\right).
\end{equation}
In turn, varying \eqref{action_wh_complex} with respect to the scalar field,
one obtains the field equation for the scalar field $\Phi$,
\begin{equation}
\label{sf_eq_gen}
\frac{1}{\sqrt{-g}}\frac{\partial}{\partial x^\mu}
\left[\sqrt{-g}g^{\mu\nu}\frac{\partial \Phi}{\partial x^\nu}\right]=
-\varepsilon \frac{d V}{d |\Phi|^2}\Phi .
\end{equation}

Our aim here is to study equilibrium wormhole solutions
and to consider their stability. For this purpose,
we take the spherically symmetric metric in the general form
 \begin{equation}
\label{metric_gen}
ds^2=e^{\nu}(dx^0)^2-e^{\lambda} dr^2-e^{\mu} d\Omega^2,
\end{equation}
where $\nu$, $\lambda$, and $\mu$ are functions of the radial coordinate $r$
and the time coordinate $x^0$
and $d\Omega^2$ is the metric on the unit two-sphere.
When considering equilibrium wormholelike configurations,
it is convenient to choose the metric in polar Gaussian coordinates
\begin{equation}
\label{metric_wh_complex}
ds^2=e^{\nu}(dx^0)^2-dr^2-R^2 d\Omega^2,
\end{equation}
where  now $\nu$ and $e^{\mu}=R^2$ are functions of $r$ only.

In the case of canonical scalar fields (i.e., when $\varepsilon=+1$),
one can obtain the well-known boson-star solutions
which possess a trivial spacetime topology.
(For a general overview on the subject of boson stars, see, e.g.,~the reviews
\cite{Jetzer:1991jr,Lee:1991ax,Schunck:2003kk,Liebling:2012fv}.)
Note that one can also obtain boson star-like objects
with a nontrivial spacetime topology, when an additional
real ghost scalar field is included
\cite{Dzhunushaliev:2014bya,Hoffmann:2017jfs}.

Here, we study the case of localized solutions
with a wormholelike topology, which is provided by the presence of
a complex ghost scalar field, i.e., for $\varepsilon=-1$ in the action~\eqref{action_wh_complex}. Then,
in order to have no time dependence in the Einstein equations,
we employ for the complex ghost scalar field
the harmonic ansatz
 \begin{equation}
\label{sf_ansatz}
\Phi(x^0,r)=\phi(r)e^{-i \omega x^0}.
\end{equation}
As in the case of boson stars, this ansatz ensures
that the spacetime of the system under consideration remains static.

The above ansatz then leads to
the following system of Einstein-scalar equations:
\begin{eqnarray}
\label{Einstein-00_complex}
&&-\left[2\frac{R^{\prime\prime}}{R}+\left(\frac{R^\prime}{R}\right)^2\right]
+\frac{1}{R^2}
={8\pi G} T_0^0=
{4\pi G} \left[ -\left(\phi^{\prime 2}+\omega^2 e^{-\nu}\phi^2\right)+V\right],
 \\
\label{Einstein-11_complex}
&&-\frac{R^\prime}{R}\left(\frac{R^\prime}{R}+\nu^\prime\right)+\frac{1}{R^2}
={8\pi G} T_1^1=
{4\pi G} \left(\phi^{\prime 2}+\omega^2 e^{-\nu}\phi^2+V\right),
\\
\label{Einstein-22_complex}
&&\frac{R^{\prime\prime}}{R}+\frac{1}{2}\frac{R^\prime}{R}\nu^\prime+
\frac{1}{2}\nu^{\prime\prime}+\frac{1}{4}\nu^{\prime 2}
=-{8\pi G} T_2^2=
-{4\pi G} \left(-\phi^{\prime 2}+\omega^2 e^{-\nu}\phi^2+V\right),\\
\label{sf_complex}
&&\phi^{\prime\prime}+\left(\frac{1}{2}\nu^\prime+2\frac{R^\prime}{R}\right)\phi^\prime+\left(\omega^2 e^{-\nu}+\frac{d V}{d |\Phi|^2}\right)\phi=0.
\end{eqnarray}
Depending on the form of the potential $V$ and the boundary conditions,
one then obtains localized equilibrium solutions by solving this
set of equations numerically.

Here, we would like to demonstrate that one can obtain wormholelike solutions
supported by a complex ghost scalar field
for a relatively simple potential containing
the quartic term $|\Phi|^4$.
Let us therefore recall that in the case of a real ghost scalar field $\Phi$
there are wormholelike regular solutions for a quartic potential of the type
$V\sim -\left(1-\bar \lambda \Phi^2\right)^2$
\cite{Kodama:1978dw,Kodama:1978zg}.
In this case, the corresponding solutions for the scalar field
start in one of the maxima of this potential
($\Phi_{\text{max}}=-1/\sqrt{\bar \lambda}$)
and end in the other maximum ($\Phi_{\text{max}}=+1/\sqrt{\bar \lambda}$).
However, when one generalizes this potential
to a complex ghost scalar field,
it is no longer possible to get regular solutions.

Hence, in order to obtain regular asymptotically flat solutions
with a nontrivial topology,
we choose another form of the quartic potential. Namely,
%following \cite{Colpi:1986ye},
we seek regular zero-node solutions
for a scalar field the potential of which has the form
\begin{equation}
\label{poten_Mex_complex}
V=-m^2 |\Phi|^2+\frac{1}{2} \bar \lambda |\Phi|^4.
\end{equation}
Here, $m$ and $\bar \lambda$ are free parameters of the scalar field.
Let us point out that this potential has also
a reversed sign in front of the quadratic term as compared to
a canonical mass term. When such a reversed sign appears for canonical
scalar fields, this signals spontaneous symmetry breaking
as in the case of the Mexican hat potential for the Higgs boson.
Together with a canonical mass term,
such a quartic self-interaction term has been used before
when modelling boson stars \cite{Colpi:1986ye}.

For the time-dependent complex ghost scalar fields,
the above potential (\ref{poten_Mex_complex})
admits only topologically trivial solutions,
which start and end at the same maximum,
in contrast to the case studied in Refs.~\cite{Kodama:1978dw,Kodama:1978zg}.
In the terminology of Lee and collaborators
\cite{Lee:1991ax,Rajaraman:1982is},
such solutions are referred to as nontopological soliton solutions,
since their existence is not based on a conserved topological current.

\subsection{Numerical solutions}
\label{equilib_confs_num_sol}

Let us now turn to the discussion of the numerical solutions.
In order to solve the above set of equations numerically, it is convenient
to introduce new dimensionless variables
\begin{equation}
\label{dmls_vars}
x=m r, \quad X=m R, \quad \Omega=\omega/m, \quad
\Lambda=\frac{\bar \lambda}{4\pi G m^2},
\quad \varphi=\sqrt{4\pi G}\phi.
\end{equation}
Then, using the potential \eqref{poten_Mex_complex},
one can rewrite Eqs.~\eqref{Einstein-00_complex}-\eqref{sf_complex} in the form
\begin{eqnarray}
\label{Einstein-00_complex_dmls}
&&-\left[2\frac{X^{\prime\prime}}{X}+\left(\frac{X^\prime}{X}\right)^2\right]+\frac{1}{X^2}
=-\varphi^{\prime 2}-\Omega^2 e^{-\nu}\varphi^2-\varphi^2+\frac{\Lambda}{2}\varphi^4,
 \\
\label{Einstein-11_complex_dmls}
&&-\frac{X^\prime}{X}\left(\frac{X^\prime}{X}+\nu^\prime\right)+\frac{1}{X^2}
=\varphi^{\prime 2}+\Omega^2 e^{-\nu}\varphi^2-\varphi^2+\frac{\Lambda}{2}\varphi^4,
\\
\label{Einstein-22_complex_dmls}
&&\frac{X^{\prime\prime}}{X}+\frac{1}{2}\frac{X^\prime}{X}\nu^\prime+
\frac{1}{2}\nu^{\prime\prime}+\frac{1}{4}\nu^{\prime 2}
=
\varphi^{\prime 2}-\Omega^2 e^{-\nu}\varphi^2+\varphi^2-\frac{\Lambda}{2}\varphi^4,\\
\label{sf_complex_dmls}
&&\varphi^{\prime\prime}+\left(\frac{1}{2}\nu^\prime+2\frac{X^\prime}{X}\right)\varphi^\prime+\left(\Omega^2 e^{-\nu}-1+\Lambda\varphi^2\right)\varphi=0.
\end{eqnarray}
 Note that these equations are not all independent because of the Bianchi identities, so one can use any three of them in calculations.
Here, we have solved the set of Eqs.~(\ref{Einstein-00_complex_dmls}),
(\ref{Einstein-22_complex_dmls}),
and (\ref{sf_complex_dmls}), treating the first-order equation
(\ref{sf_complex_dmls}) as a constraint equation to monitor
the accuracy of the results, since it ought to be satisfied
identically for any solution of the chosen set of equations.

\begin{figure}[t]
\centering
  \includegraphics[height=18cm]{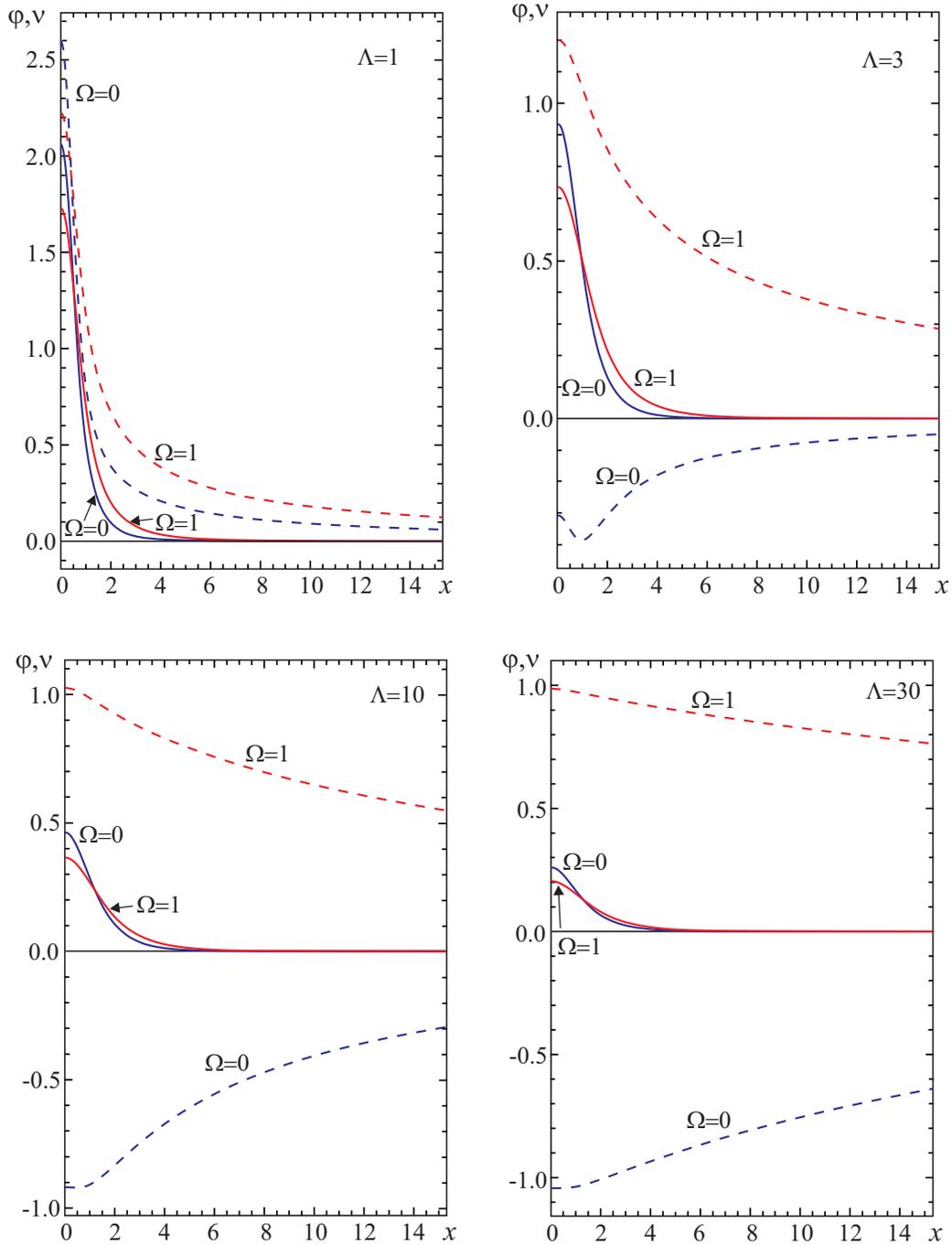}
%\vspace{-1.cm}
\caption{Typical solutions for the scalar field function $\varphi$ (solid lines)
and the metric function $\nu$ (dashed lines) for different values of
self-coupling constant $\Lambda$.
(In view of the symmetry $x\to -x$, only the solutions for positive $x$ are shown.)
For the boson frequency $\Omega$, the lower limit $\Omega=0$ is chosen,
while the upper limit is represented by the value $\Omega=1$
in all graphs.
}
\label{fig_nu_sigma}
\end{figure}

When solving the above set of equations,
we use the following symmetric boundary conditions given
at the center $r=0$,
%these are given by
\begin{equation}
\label{bc_compl}
X(0)=X_c, \quad \nu(0)=\nu_c, \quad \varphi(0)=\varphi_c,
\end{equation}
with all first-order derivatives being equal to zero at the center.
Then, the constraint equation~\eqref{Einstein-11_complex_dmls} yields
\begin{equation}
\label{throat_radius}
X_c=\frac{1}{\varphi_c\sqrt{\Omega^2 e^{-\nu_c}-1+(\Lambda/2)\varphi_c^2}}.
 \end{equation}
Taking this expression into account and expanding the metric function
$X$ in the vicinity of the center as $X\approx X_c+1/2\, X_2 x^2$,
one finds from Eq.~\eqref{Einstein-00_complex_dmls}
the following value for the second derivative of $X$ at the center
$$
X_2=\Omega^2 e^{-\nu_c}\varphi_c^2 X_c,
$$
which is always positive.
This means that the quantity $X_c$ corresponds to the radius
of the wormhole throat (and not to the radius of an equator)
residing at the center, since the  wormhole throat
is defined by $X_{\text{th}}=\text{min}\{X(x)\}$.
Consequently, throughout the paper, we will deal only with configurations
possessing a single wormhole throat.

So, for the systems under consideration,
we have three parameters, $\varphi_c$, $\nu_c$, and $\Omega$,
one of which can be chosen freely. For example, as in the case
of boson stars \cite{Colpi:1986ye},
this free parameter can be chosen to be $\varphi_c$.
Then, the other two parameters must be chosen in such a way
as to provide asymptotic flatness of the spacetime,
when $\varphi$, $\varphi^\prime$, and $\nu \to 0$ and $X\to x$.
In this sense, we are dealing with an eigenvalue problem for
the parameters $\nu_c$ and $\Omega$.
The corresponding asymptotic behavior of the solutions is as follows,
\begin{eqnarray}
\label{asymp_sol}
&&e^{\nu}\to 1-\frac{2 C_2}{x}, \quad X \to x,
\quad X^\prime \to 1-\frac{C_2}{x}, \nonumber \\
&&\varphi \to C_1  \exp{\left(-\sqrt{1-\Omega^2}\,x\right)} x^\beta \quad \text{for} \quad 0 \leqslant \Omega < 1, \quad
\varphi \to C_3\frac{\exp{\left(-\sqrt{8 |C_2|x}\right)}}{x^{3/4}} \quad\text{for} \quad  \Omega=1,
\end{eqnarray}
where $C_1, C_2$, and $C_3$ are integration constants and $\beta=-1+C_2 \Omega^2/\sqrt{1-\Omega^2}$.
Note that at $\Omega\to 1$ the integration constant $C_2$,
which corresponds to the Arnowitt–Deser–Misner  (ADM) mass of the wormhole,
is always negative
for the configurations under consideration.

We have solved the set of
equations~\eqref{Einstein-00_complex_dmls}-\eqref{sf_complex_dmls} numerically,
together with the boundary conditions \eqref{bc_compl} and \eqref{throat_radius},
varying the strength $\Lambda$ of the self-interaction
in the interval $1 \leqslant \Lambda \leqslant 20000$
and the value of the boson frequency $\Omega$
in the interval $0 \leqslant \Omega \leqslant 1$.
We demonstrate a set of typical solutions for the scalar field
function $\varphi$ and the metric function $\nu$ in Fig.~\ref{fig_nu_sigma},
where we display the solutions only in one asymptotically flat part
of the spacetime, since the solutions are symmetric with respect to~$x \to -x$.
Here, four different values of $\Lambda$ are chosen,
$\Lambda=1$, 3, 10, and 30.

The values of the boson frequency chosen in the plots,
$\Omega=0$ and $\Omega = 1$, %(i.e., here $\Omega=0.999$),
represent the boundary values of the physically acceptable interval.
Note that the limiting case $\Omega=0$ corresponds to configurations
supported by a real ghost scalar field.
For values of $\Omega$ in between these limiting values,
the graphs of the above functions will lie between
those shown in Fig.~\ref{fig_nu_sigma}.

\begin{figure}[t]
\begin{minipage}[t]{.49\linewidth}
  \begin{center}
  \includegraphics[width=8.5cm]{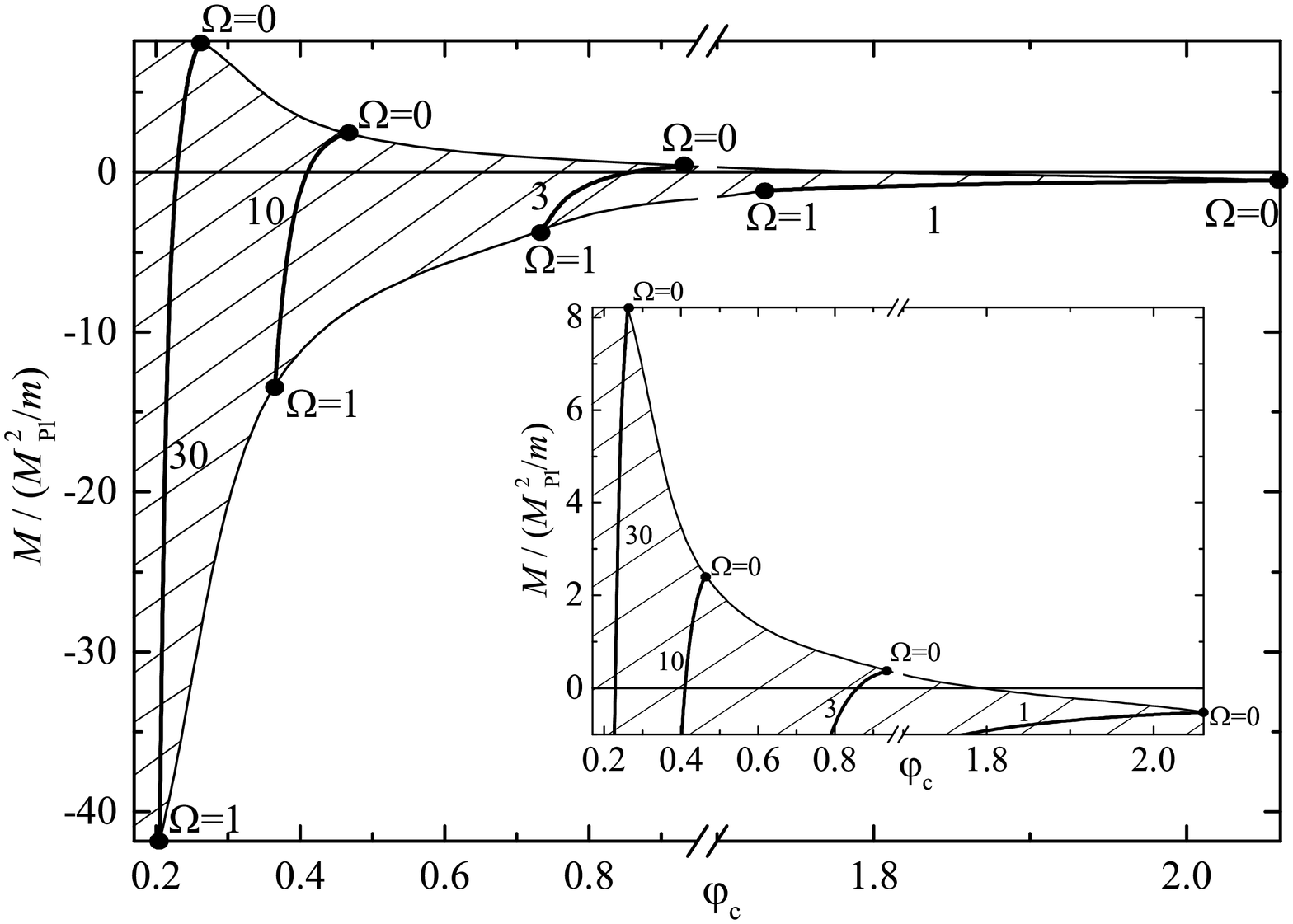}
  \end{center}
\end{minipage}\hfill
\begin{minipage}[t]{.49\linewidth}
  \begin{center}
  \includegraphics[width=8.75cm]{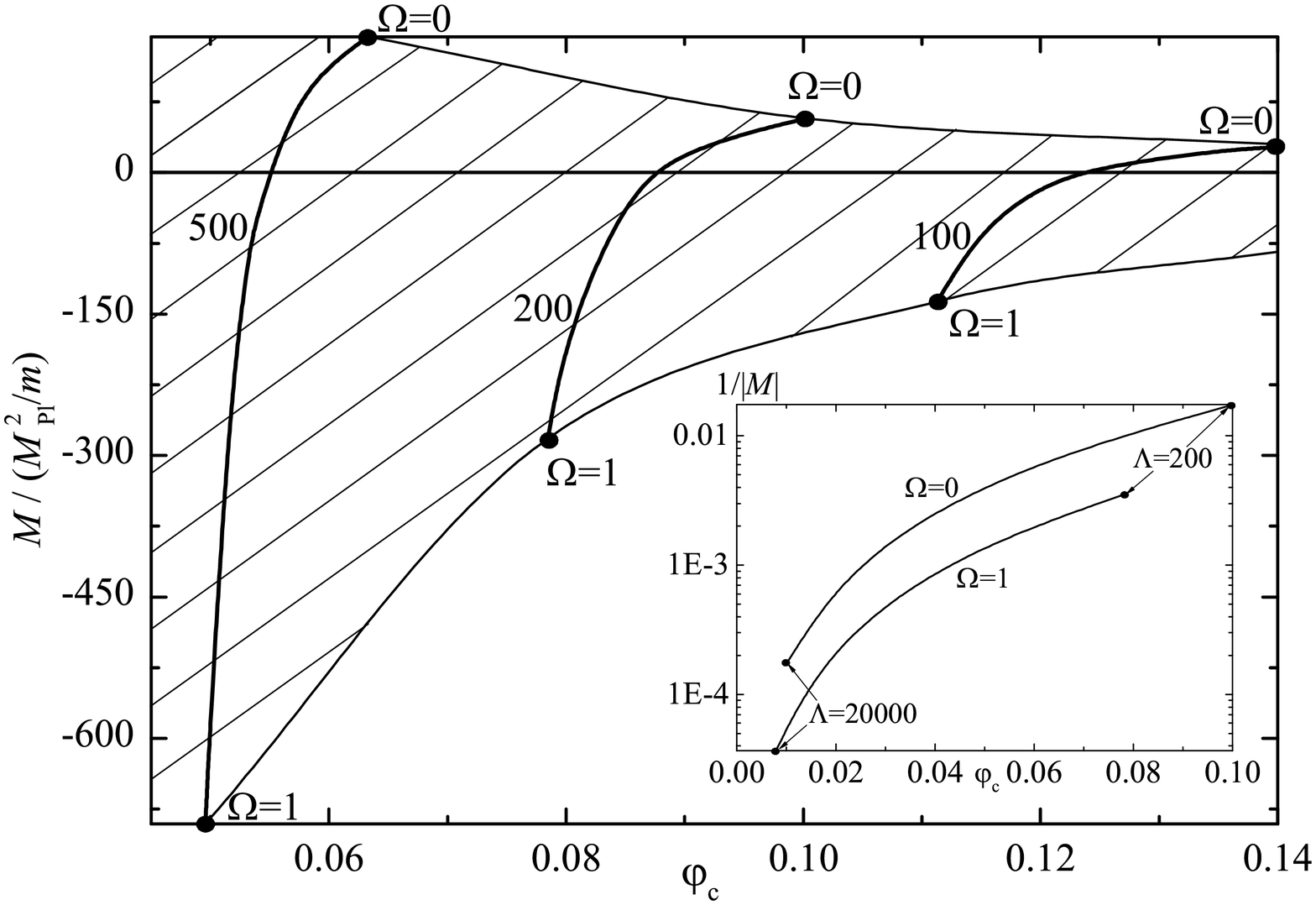}
  \end{center}
\end{minipage}\hfill
\vspace{-0.3cm}
\caption{
The bold curves show the wormhole mass $M$ as a function
of the central value of the scalar field $\varphi_c$ for different values of
the self-coupling constant $\Lambda$ (designated by the numbers near the curves).
For other values of $\Lambda$, the masses lie within the shaded regions,
enclosed by the limiting curves for the boson frequencies
$\Omega=0$ and $\Omega=1$.
In the left panel, the inset shows the region
where the masses are close to zero.
In the right panel, the inset demonstrates the growth
of the modulus of the  ADM mass as $\phi_c \to 0$,
when $\Lambda\to \infty$.
}
\label{fig_M_sigma}
\end{figure}

\begin{figure}[h!]
\begin{minipage}[t]{.49\linewidth}
  \begin{center}
  \includegraphics[width=8.5cm]{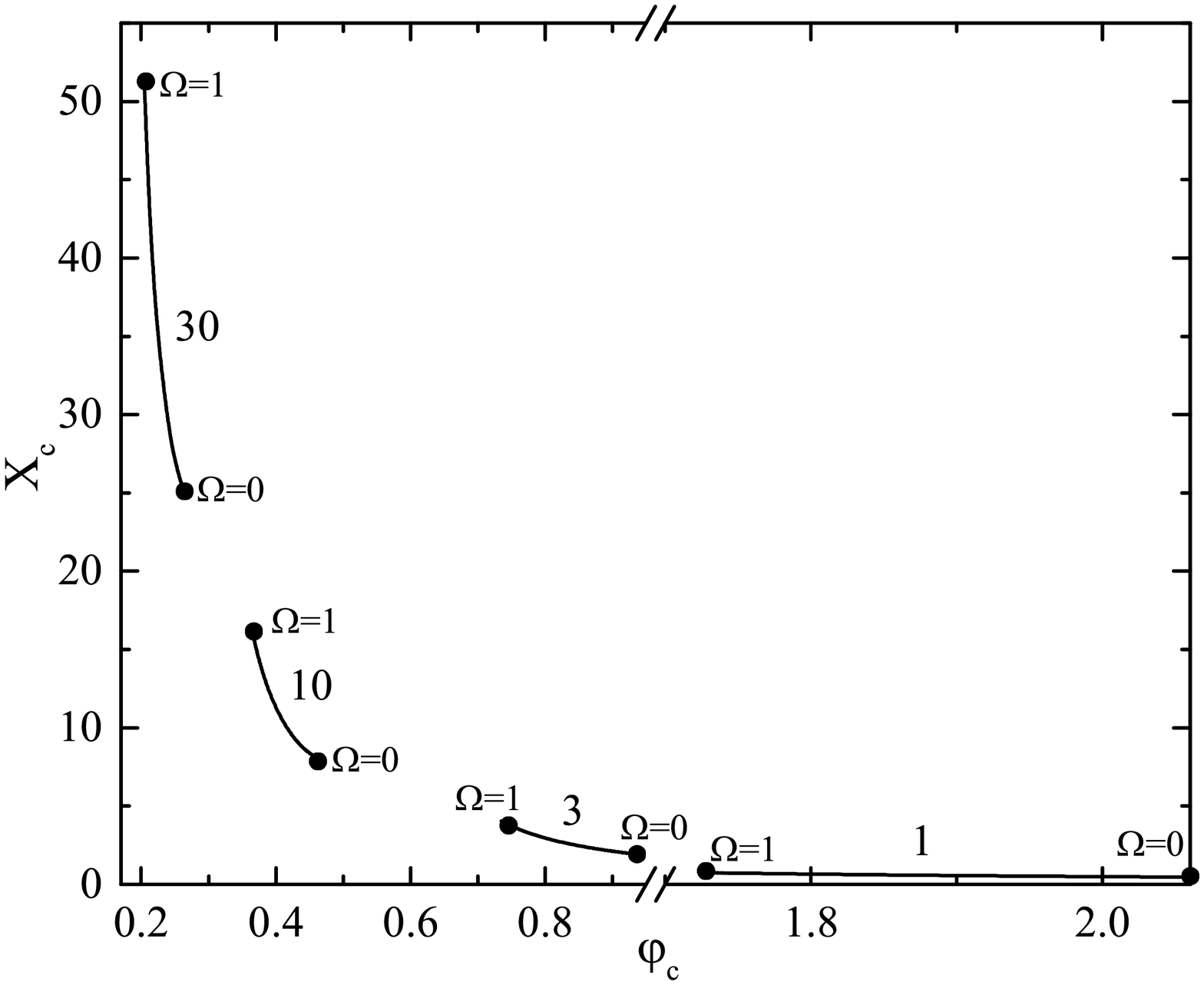}
  \end{center}
\end{minipage}\hfill
\begin{minipage}[t]{.49\linewidth}
  \begin{center}
  \includegraphics[width=8.5cm]{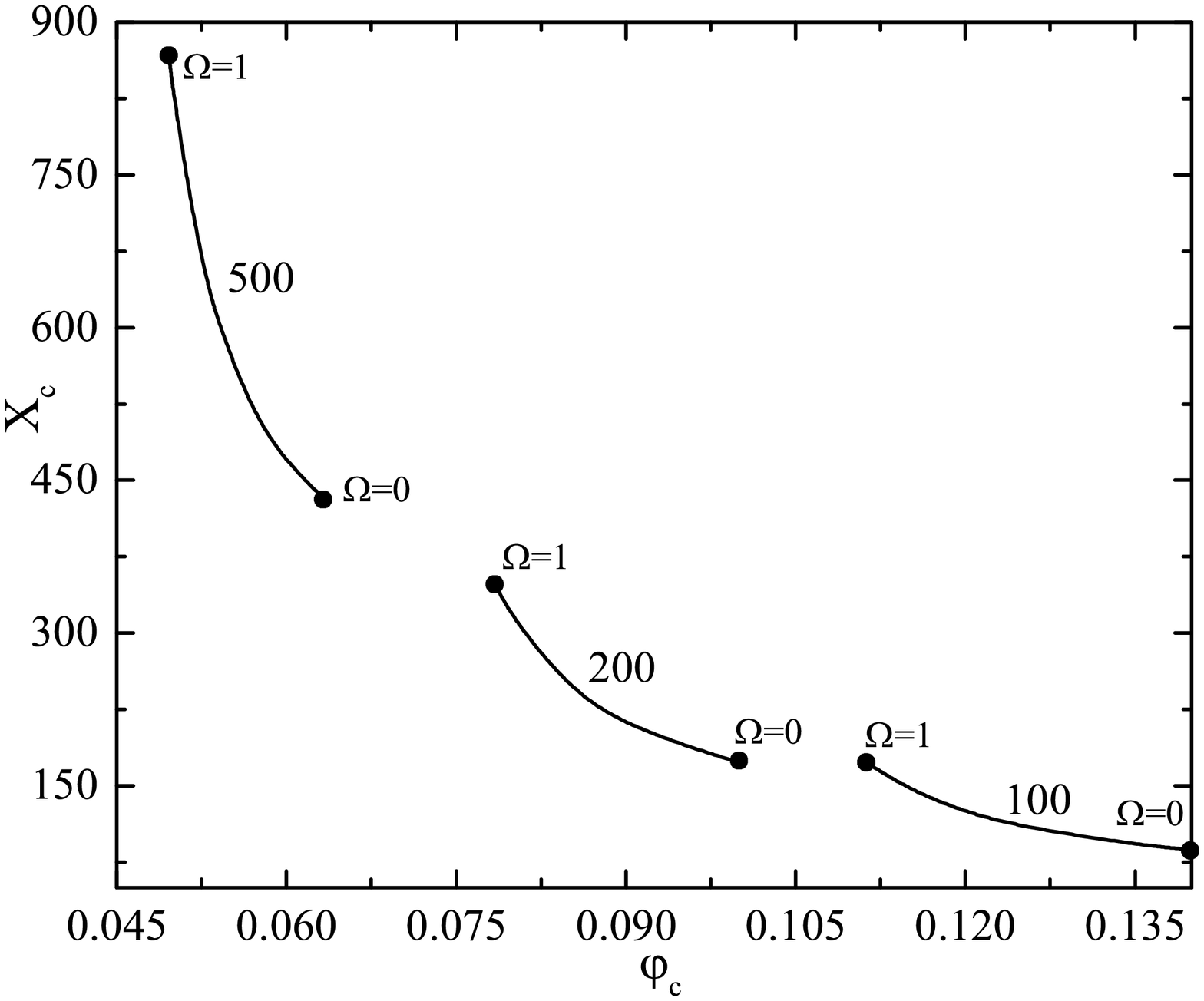}
  \end{center}
\end{minipage}\hfill
\vspace{-0.3cm}
\caption{
The dimensionless radius $X_c$ \eqref{throat_radius} of the wormhole throat
is shown as a function of the central value of the scalar field $\varphi_c$
for different values of
the self-coupling constant $\Lambda$ (designated by the numbers near the curves).
The end points of the curves correspond to the boson frequencies
$\Omega=0$ and $\Omega=1$.
}
\label{fig_Sigma_c_sigma}
\end{figure}

Let us now consider the  ADM mass of the above systems.
In the case of spherical symmetry,
the  Misner-Sharp
\cite{Misner:1964je} mass $M(r)$ associated with the volume
enclosed by a sphere with
circumferential radius $R_c$, corresponding to the center
of the configuration, and another sphere with
circumferential radius $R>R_c$
can be defined as follows:
\begin{equation}
\label{mass_dm}
M(r)=\frac{1}{2 G}R_c+{4\pi} \int_{R_c}^{r} T_0^0 R^2   dR.
\end{equation}
Taking the boundary to (spacelike) infinity,
the Misner-Sharp mass leads to the ADM mass.
In the dimensionless variables \eqref{dmls_vars}, this becomes
\begin{equation}
\label{mass_dmls}
{\cal M}(x)\equiv\frac{M(x)}{M^2_{\text{Pl}}/m}=
\frac{1}{2}\left[X_c-\int_0^x \left(
\varphi^{\prime 2}+\Omega^2 e^{-\nu} \varphi^2+\varphi^2-\frac{\Lambda}{2}\varphi^4
\right)X^2\frac{dX}{d x^\prime}d x^\prime\right].
\end{equation}

The results of the calculations of the mass are shown in Fig.~\ref{fig_M_sigma}.
It is interesting to compare these results with those obtained for systems
with a trivial spacetime topology -- boson stars
(see, e.g., Refs.~\cite{Jetzer:1991jr,Lee:1991ax,Schunck:2003kk,Liebling:2012fv}).
For the boson stars, solutions are sought for
some range of central values of the scalar field $\varphi_c$,
starting from zero, which leads to a set of configurations with different masses.
The typical behavior of the dependence of the mass on the central
value of the scalar field is as follows:
As $\varphi_c \to 0$, the mass of the system goes to zero as well,
and the boson frequency goes to its upper limit, $\Omega \to 1$.
When $\varphi_c$ increases, the mass at first also increases,
reaching the maximum value at some critical $\varphi_c^{\text{cr}}$.
Then, the mass decreases, reaches a local minimum, increases again,
and exhibits a damped oscillation toward some limiting value
at large $\varphi_c$, associated with a limiting
boson frequency $\Omega\neq 0$.

In contrast to those systems with a trivial topology,
the configurations with a nontrivial topology considered here
possess the following distinctive features:
\vspace{-0.2cm}
\begin{enumerate}
\itemsep=-0.2pt
\item[(1)] %1.
When $\varphi_c$ increases, there is a monotonic change of the mass (no extrema).
 \item[(2)] Regular solutions exist
for all physically acceptable values $0\leqslant \Omega \leqslant 1$.
\item[(3)] As $\Lambda \to \infty$,
the central value of the scalar field $\varphi_c \to 0$.
In this case, both the size of the throat
$X_c$ and the  ADM mass of the system increase without limit,
as shown for the mass in the inset of the right panel of Fig.~\ref{fig_M_sigma}.
(Note that the mass is positive for $\Omega \sim 0$ and large $\Lambda$
and negative for $\Omega \to 1$.)
\end{enumerate}

Figure~\ref{fig_Sigma_c_sigma} demonstrates how the dimensionless radius
$X_c$ of the wormhole throat, which also enters the
equation for the mass \eqref{mass_dmls},
changes with the self-coupling constant $\Lambda$.
Comparing Figs.~\ref{fig_M_sigma} and \ref{fig_Sigma_c_sigma},
one notes that, despite the fact that with increasing $\Omega$
the radius of the throat increases, the  ADM mass of the system decreases.

\begin{figure}[t]
\begin{minipage}[t]{.49\linewidth}
  \begin{center}
  \includegraphics[width=8.5cm]{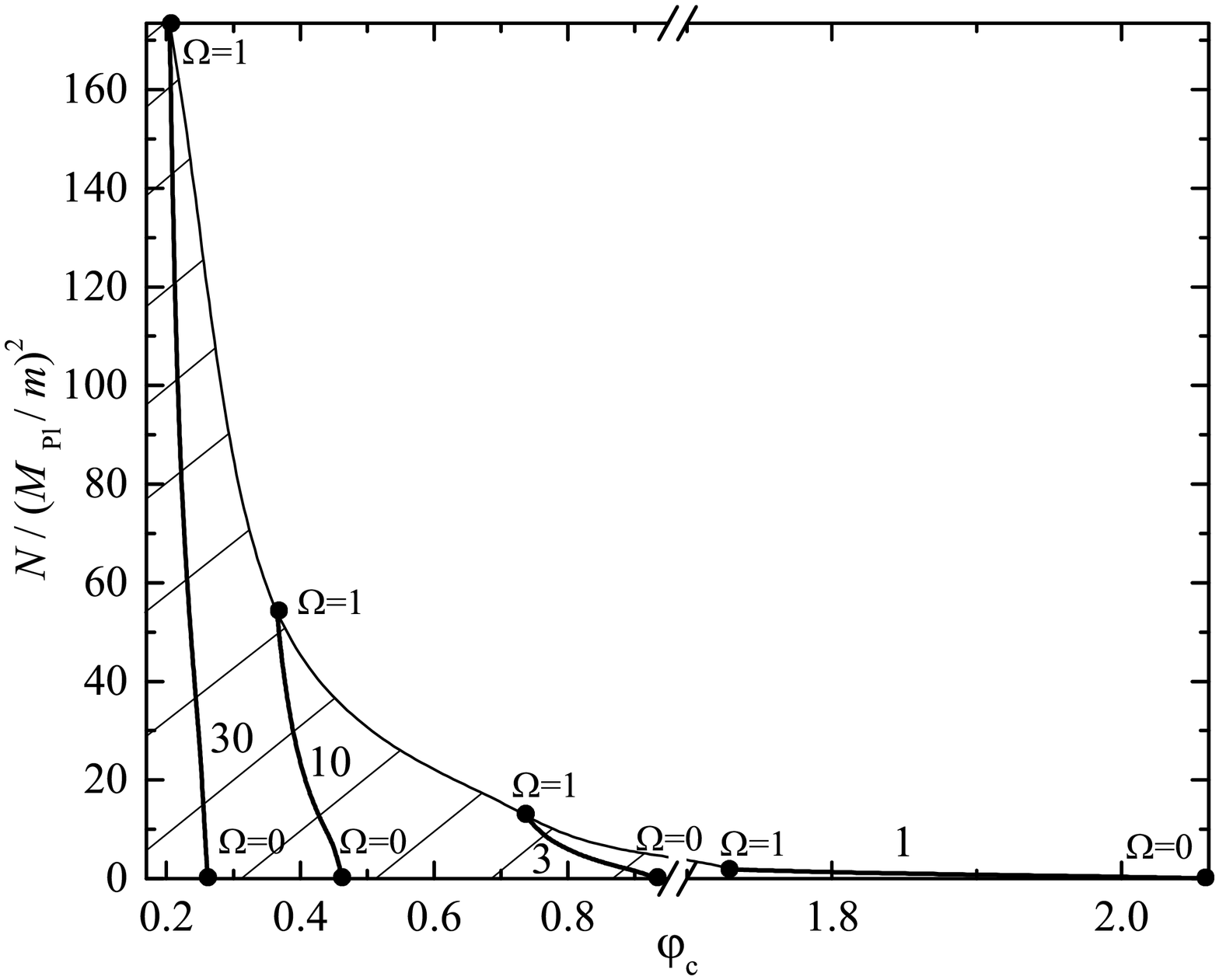}
    \end{center}
\end{minipage}\hfill
\begin{minipage}[t]{.49\linewidth}
  \begin{center}
  \includegraphics[width=8.5cm]{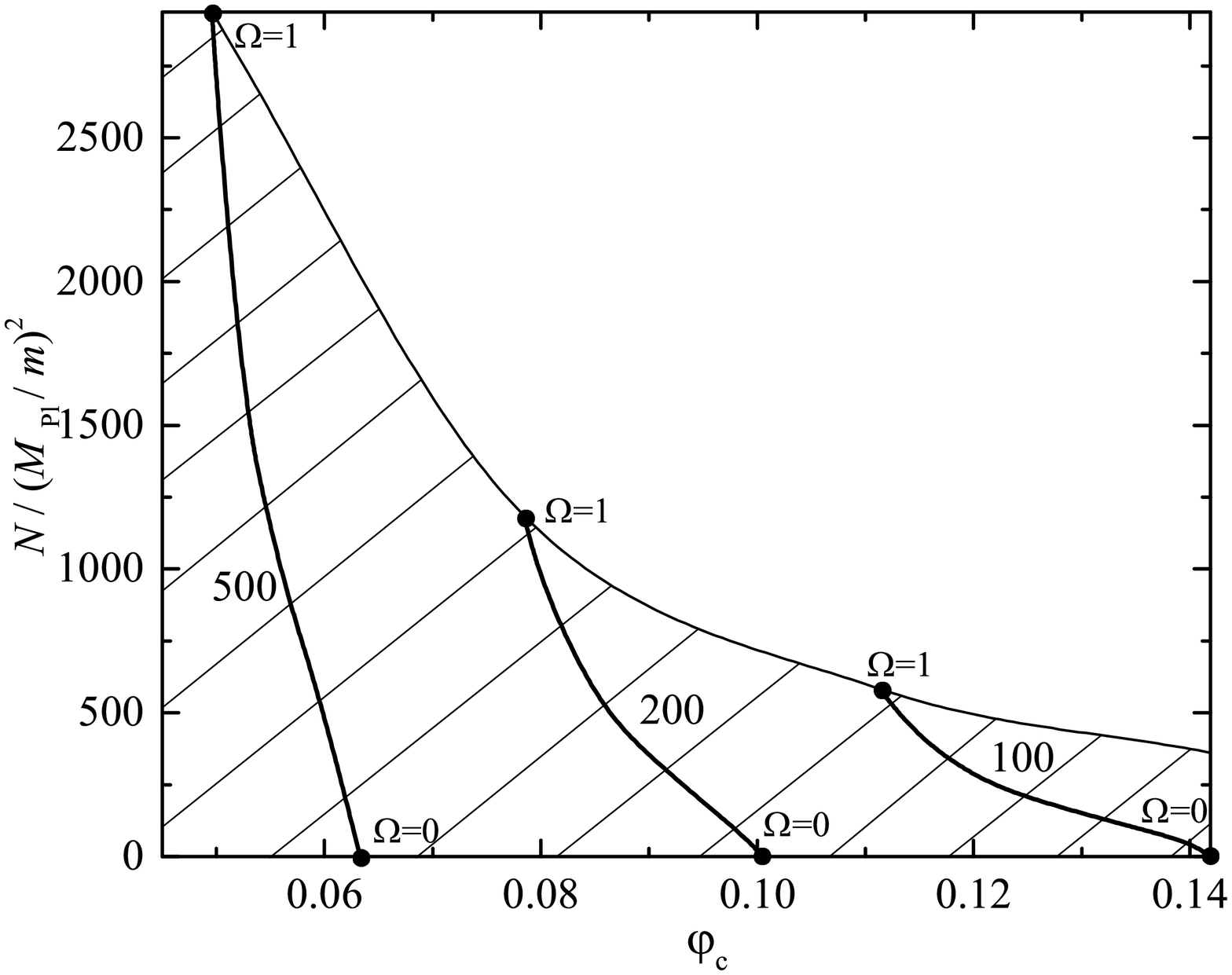}
  \end{center}
\end{minipage}\hfill
\vspace{-0.3cm}
\caption{
The number of particles $N$ as a function of the central value
of the scalar field $\varphi_c$ for different values of
the self-coupling constant $\Lambda$ (designated by the numbers near the curves).
For other values of $\Lambda$, the number of particles lies
within the shaded regions,
enclosed by the limiting curves for the boson frequencies
$\Omega=0$ and $\Omega=1$.
}
\label{fig_N_sigma}
\end{figure}

Let us finally calculate the total number of particles $N$ forming the system.
$N$ can be obtained from the continuity equation $j^\mu_{;\mu}=0$,
where the four-current $j^\mu$ is given by
$$
j^\mu=i g^{\mu\nu}\left(\partial_\nu\Phi \Phi^*-\partial_\nu\Phi^* \Phi\right).
$$
This leads to the conserved charge
\begin{equation}
\label{tot_part_number}
N=\int \sqrt{-g} j^0 d^3 x=8\pi \omega \int_0^r  e^{-\nu/2} R^2 \phi^2 dr^\prime=
\frac{2 \Omega}{\left(m/M_\text{Pl}\right)^2}\int_0^x  e^{-\nu/2} X^2 \varphi^2 dx^\prime.
\end{equation}
The results for the total particle number are shown in Fig.~\ref{fig_N_sigma}.
It is seen that for a fixed finite value of the boson frequency $\Omega$
the number of particles increases with increasing $\Lambda$.
Note that the total particle number vanishes in the limit
$\Omega=0$. This corresponds to the fact that for a real scalar field %(i.e. when $\Omega=0$)
%the number of particles is equal to the number of antiparticles.
there is no such conserved current.

\section{Linear stability analysis}
\label{stab_analysis}

In this section, we perform a linear stability analysis of
the above equilibrium solutions.
We derive the appropriate set of equations and then
analyze them numerically.
In doing so, we start from the stability analysis of the systems
supported by a real scalar field;
i.e., we first consider the case of  $\Omega=0$.
Our aim here is to clarify how the generalization of these configurations
to the case of a complex scalar field with $\Omega\ne 0$
influences their stability.

In performing the stability analysis, we start from the general metric
\eqref{metric_gen}.
The perturbed solutions are then sought in the form
$$
\nu=\nu_0(r)+\nu_p(r,x^0), \quad \lambda=\lambda_0(r)+\lambda_p(r,x^0), \quad
\mu=\mu_0(r)+\mu_p(r,x^0).
$$
The subscript 0 denotes the static background solutions from the previous section,
and the subscript $p$ denotes the perturbations.
Also, we represent the complex scalar field
%in terms of two real fields, $\Phi_1$ and $\Phi_2$,
in terms of two distinct  real functions, $\Phi_1$ and $\Phi_2$,
retaining the harmonic time dependence of the unperturbed solutions
as a factor
$$
\Phi(r, x^0)=\left[\Phi_1(r, x^0)+i \Phi_2(r,x^0)\right]e^{-i \omega x^0}.
$$
For the unperturbed system, $\Phi_1(r,x^0)\to \phi_0(r)$, and $\Phi_2=0$.
The perturbative expressions for these scalar functions
are then taken in the form
$$
\Phi_1=\phi_0(r)+\Phi_{1p}(r,x^0), \quad \Phi_2= \Phi_{2p}(r,x^0).
$$

Using the above expressions,
one can get the following %$(^0_0), (^1_1), (^2_2)$
perturbed components of the Einstein equations
$E^0_0$, $E^1_1$, and $E^2_2$
(here and below, we take $\lambda_0=0$,
since in obtaining the equilibrium solutions
in Sec.~\ref{equilib_confs} we have employed this gauge),
\begin{eqnarray}
\label{Einstein-00pert}
&&
\mu_p^{\prime\prime}+\frac{3}{2}\mu_0^\prime \mu_p^\prime-
\frac{1}{2}\mu_0^\prime\lambda_p^\prime-
\lambda_p\left(\mu_0^{\prime\prime}+\frac{3}{4}\mu_0^{\prime 2}\right)
+e^{-\mu_0}\mu_p \nonumber \\
&&=-4\pi G
\left\{
-\phi_0 e^{-\nu_0}\omega\left[2\left(\omega \Phi_{1p}-\dot{\Phi}_{2p}\right)-\omega \phi_0\nu_p\right]-
\phi_0^\prime \left(2\Phi_{1p}^\prime-\phi_0^\prime \lambda_p\right)+V_p
\right\}
,\\
\label{Einstein-11pert}
&&
\frac{1}{2}\left[
 \left(\nu_p^\prime+\mu_p^\prime\right)\mu_0^\prime+\nu_0^\prime \mu_p^\prime-
 \lambda_p\left(\frac{1}{2}\mu_0^{\prime 2}+\mu_0^\prime \nu_0^\prime\right)
\right]-e^{-\nu_0}\ddot{\mu}_p+e^{-\mu_0}\mu_p \nonumber \\
&&=-4\pi G
\left\{
\phi_0 e^{-\nu_0}\omega\left[2\left(\omega \Phi_{1p}-\dot{\Phi}_{2p}\right)-\omega \phi_0\nu_p\right]+
\phi_0^\prime \left(2\Phi_{1p}^\prime-\phi_0^\prime \lambda_p\right)+V_p
\right\}
,\\
\label{Einstein-22pert}
&&
\mu_p^{\prime\prime}+\nu_p^{\prime\prime}+\mu_0^\prime\left(\mu_p^\prime+\frac{1}{2}\nu_p^\prime-\frac{1}{2}\lambda_p^\prime\right)+
\nu_0^\prime\left(\frac{1}{2}\mu_p^\prime-\frac{1}{2}\lambda_p^\prime+\nu_p^\prime\right)
-\lambda_p\left[\mu_0^{\prime\prime}+\nu_0^{\prime\prime}+
\frac{1}{2}\left(\mu_0^{\prime 2}+\nu_0^{\prime 2}+\mu_0^\prime \nu_0^\prime\right)\right]
\nonumber \\
&&-e^{-\nu_0}(\ddot{\mu_p}+\ddot{\nu_p})=
-8\pi G
\left\{
\phi_0 e^{-\nu_0}\omega\left[2\left(\omega \Phi_{1p}-\dot{\Phi}_{2p}\right)-\omega \phi_0\nu_p\right]-
\phi_0^\prime \left(2\Phi_{1p}^\prime-\phi_0^\prime \lambda_p\right)+V_p
\right\},
\end{eqnarray}
where $V_p$ is the perturbation of the potential.
Next, from the $(^1_0)$  component of the Einstein
equations, we have to linear order
\begin{equation}
\label{Einstein-10pert}
2\dot{\mu_p}^\prime-\dot{\lambda_p}\mu_0^\prime
+\dot{\mu_p}\left(\mu_0^\prime-\nu_0^\prime\right)
=16\pi G\left[
\phi_0^\prime \dot{\Phi}_{1p}-\omega \phi_0 \Phi_{2p}^\prime
+\omega \phi_0^\prime \Phi_{2p}
\right].
\end{equation}

In turn, for the components of the scalar field,
there are two equations (coming from the real and imaginary parts),
\begin{eqnarray}
\label{sf-real-pert}
&& \Phi_{1p}^{\prime\prime}+
\left(\frac{1}{2}\nu_0^\prime+\mu_0^\prime\right)\Phi_{1p}^{\prime}
-e^{-\nu_0} \left(\ddot{\Phi}_{1p}+2\omega \dot{\Phi}_{2p}\right)-\omega^2 e^{-\nu_0}  \phi_0 \nu_p+
 \phi_0\left[\omega^2 e^{-\nu_0} -m^2+\bar \lambda \phi_0^2\right]\lambda_p\nonumber\\
&&+\left(\omega^2 e^{-\nu_0}-m^2+3\bar \lambda\phi_0^2 \right)\Phi_{1p}
+
\frac{1}{2}\phi_0^\prime\left(\nu_p^\prime-\lambda_p^\prime+2\mu_p^\prime\right)=0,\\
&&\Phi_{2p}^{\prime\prime}+\left(\frac{1}{2}\nu_0^\prime+\mu_0^\prime\right)\Phi_{2p}^{\prime}-
\frac{1}{2}\left(\dot{\nu}_p-\dot{\lambda}_p-2\dot{\mu}_p\right)\omega e^{-\nu_0}\phi_0-e^{-\nu_0}\left(\ddot{\Phi}_{2p}-2\omega \dot{\Phi}_{1p}\right)
\nonumber\\
&&+
\left(-m^2+\omega^2 e^{-\nu_0}+\bar\lambda \phi_0^2\right)\Phi_{2p}=0,
\label{sf-im-pert}
\end{eqnarray}
where $\bar \lambda$ is the self-coupling constant from
Eq.~\eqref{poten_Mex_complex}.
Note here that combining Eqs.~\eqref{Einstein-00pert}-\eqref{sf-real-pert} and letting
$\Phi_{2p}=\omega=m=\bar\lambda=0$ one can recover  the
perturbed equations for  the real massless scalar field
of Ref.~\cite{Gonzalez:2008wd}
and perform the corresponding calculations for  the perturbations.

So far, in this section, we have used the gauge freedom
to choose the radial coordinate $r$ such that
we may set $\lambda_0=0$.
We have another gauge freedom left concerning the choice
of the metric perturbations $\nu_p, \lambda_p$, and $\mu_p$.
Here, we  take the gauge $\nu_p=\lambda_p-2\mu_p$ \cite{Dzhunushaliev:2013lna},
which permits us to exclude $\nu_p$ and to simplify the equations
accordingly.
Thus, we are left with six equations \eqref{Einstein-00pert}-\eqref{sf-im-pert}
for four functions, $\lambda_p, \mu_p, \Phi_{1p}$, and $\Phi_{2p}$,
two of which are first-order equations (constraints).

In solving this system of equations, one can proceed as follows
\cite{Gleiser:1988rq,Gleiser:1988ih}.
To reduce the number of equations, one can get rid of the function $\Phi_{2p}$.
To do this, we exclude the term $\dot{\Phi}_{2p}$ from
\eqref{sf-real-pert} by expressing $\dot{\Phi}_{2p}$
from any of  Eqs.~\eqref{Einstein-00pert}-\eqref{Einstein-22pert}.
For example, from Eq.~\eqref{Einstein-00pert}, we have
\begin{eqnarray}
\label{dotPhi2p}
&&\dot{\Phi}_{2p}=\omega\left[\Phi_{1p}-\phi_0\left(\frac{1}{2}\lambda_p-\mu_p\right)\right]-\frac{1}{2\phi_0 e^{-\nu_0} \omega} \nonumber\\
&&\times\left\{
-\phi_0^\prime \left(2\Phi_{1p}^\prime-\phi_0^\prime \lambda_p\right)+V_p+
\frac{1}{4\pi G}\left[\mu_p^{\prime\prime}+\frac{3}{2}\mu_0^\prime \mu_p^\prime-
\frac{1}{2}\mu_0^\prime\lambda_p^\prime-
\lambda_p\left(\mu_0^{\prime\prime}+\frac{3}{4}\mu_0^{\prime 2}\right)
+e^{-\mu_0}\mu_p\right]
\right\}.
\end{eqnarray}
In turn, taking sums of the components of the Einstein equations
\eqref{Einstein-00pert}+\eqref{Einstein-11pert}
and \eqref{Einstein-00pert}+\eqref{Einstein-22pert},
one can get rid of $\dot{\Phi}_{2p}$ in these equations as well.

To proceed with the stability analysis,
we now assume that the perturbations have the following time dependence,
\begin{equation}
\label{harmonic}
y_p(x^0, r) = \bar{y}_p(r) e^{i\chi x^0}~,
\end{equation}
where $y$ is any of the functions $\mu$, $\lambda$, and $\Phi_{1}$
and the functions $\bar{y}_p(r)$ depend only on the spatial coordinate $r$.
For convenience, we hereafter drop the bar.
Then, we get the following set of gravitational equations,
 \begin{eqnarray}
\label{00plus11-pert}
 \text{Eq.}~\eqref{Einstein-00pert}+\text{Eq.}~\eqref{Einstein-11pert}:\quad&&
\mu_p^{\prime\prime}+\left(\frac{1}{2}\nu_0^\prime
+\mu_0^\prime\right)\mu_p^\prime-\left( \mu_0^{\prime\prime}
+\mu_0^{\prime 2}+\frac{1}{2}\mu_0^{\prime}\nu_0^{\prime}\right)\lambda_p+
\left(2 e^{-\mu_0}+\chi^2 e^{-\nu_0}\right)\mu_p=-8\pi G V_p,\\
  \text{Eq.}~\eqref{Einstein-00pert}+\text{Eq.}~\eqref{Einstein-22pert}:\quad&&
 \mu_p^{\prime\prime}+\lambda_p^{\prime\prime}
+\left(\mu_0^\prime-\frac{1}{2}\nu_0^{\prime}\right)
\left(3 \mu_p^{\prime}-\lambda_p^{\prime}\right)-
 \left[3\mu_0^{\prime\prime}+\nu_0^{\prime\prime}+\frac{1}{2}\left(
 4\mu_0^{\prime 2}+\nu_0^{\prime 2}
+\mu_0^{\prime} \nu_0^{\prime} \right)\right]\lambda_p
  \nonumber\\
 &&+\left(2 e^{-\mu_0}+\chi^2 e^{-\nu_0}\right)\mu_p
+\chi^2 e^{-\nu_0}\lambda_p=-16\pi G\left[
 -\phi_0^\prime \left(2\Phi_{1p}^\prime-\phi_0^\prime \lambda_p\right)
+V_p \right],
 \label{00plus22-pert}
\end{eqnarray}
where the perturbation of the potential is
$V_p=2\phi_0\left(-m^2+\bar\lambda \phi_0^2\right)\Phi_{1p}$.
These two equations are supplemented by the equation
for the scalar field \eqref{sf-real-pert} with \eqref{dotPhi2p}.
As a result, we obtain the system of three Eqs.~\eqref{sf-real-pert},
\eqref{00plus11-pert}, and \eqref{00plus22-pert}
for the three functions $\mu_p, \lambda_p$, and $\Phi_{1p}$.

For this set of equations, we choose the following boundary conditions
at $r=0$,
\begin{equation}
\label{bound_cond_pert_compl}
\lambda_p(0)=\lambda_{p 0}, \quad
\mu_p(0)=\mu_{p 0},\quad
\Phi_{1p}(0)=\Phi_{1p0},
\end{equation}
where all functions are even functions.

\begin{figure}[t]
\begin{minipage}[t]{.49\linewidth}
  \begin{center}
  \includegraphics[width=8.5cm]{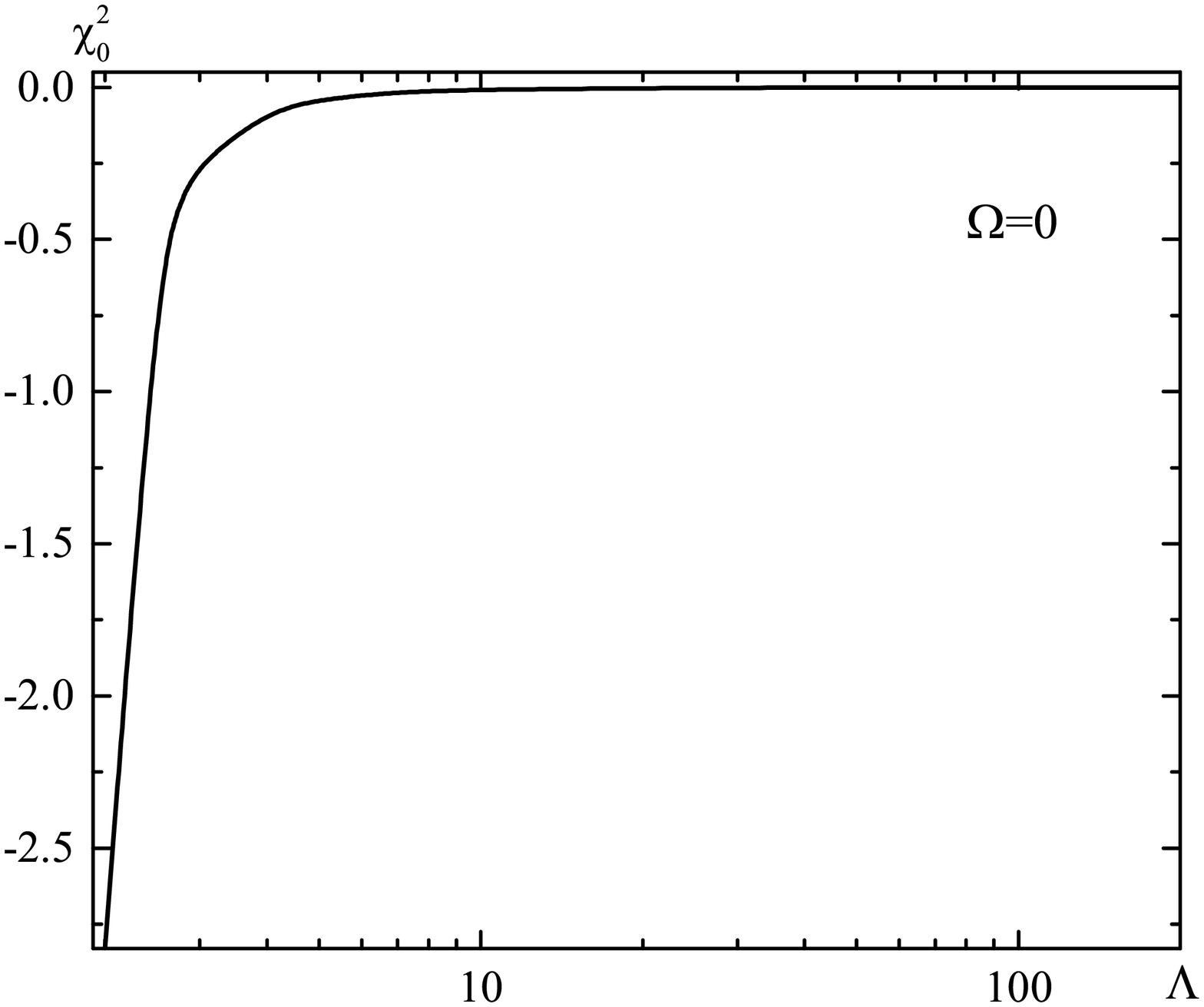}
    \end{center}
\end{minipage}\hfill
\begin{minipage}[t]{.49\linewidth}
  \begin{center}
  \includegraphics[width=8.5cm]{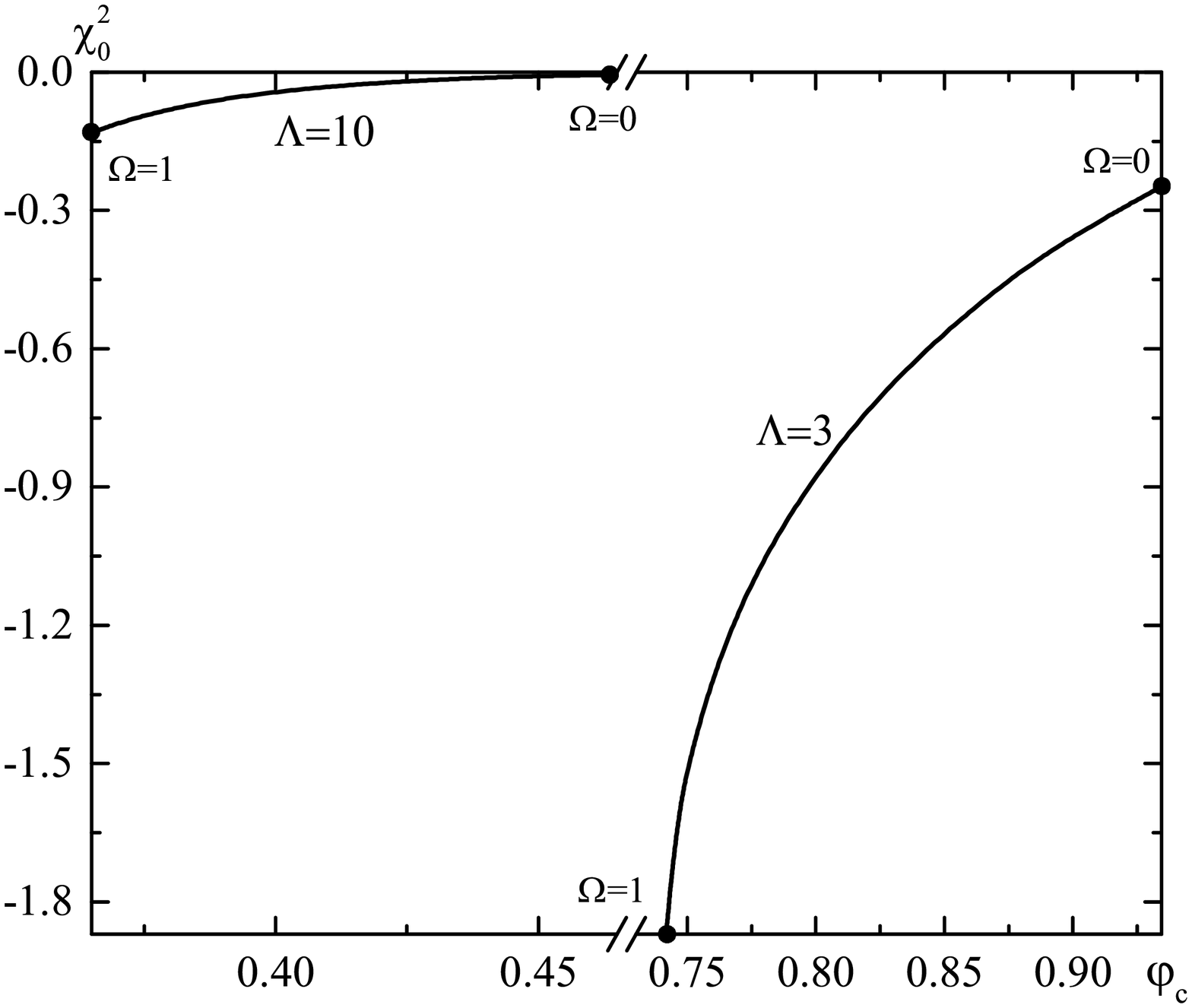}
  \end{center}
\end{minipage}\hfill
\vspace{-0.3cm}
\caption{Left panel:
The lowest eigenvalue $\chi_0^2$ is shown as a function of
the self-coupling constant $\Lambda$ for
the case of a real scalar field (with $\Omega=0$ and $\Phi_2=0$).
For large $\Lambda$, $\chi_0^2 \to -0$.
Right panel: The lowest eigenvalue $\chi_0^2$ is shown as a function
of the central value of the scalar field $\varphi_c$
for two values of the self-coupling constant $\Lambda$
in the physically acceptable interval of the boson frequency
$0 \leqslant \Omega \leqslant 1$.
}
\label{fig_chi}
\end{figure}

Using the dimensionless variables \eqref{dmls_vars}
and introducing also the dimensionless perturbation
$\Phi_{1p}\to\sqrt{4\pi G} \Phi_{1p}$,
we can now search for numerical solutions to the set of
Eqs.~\eqref{sf-real-pert}, \eqref{00plus11-pert}, and \eqref{00plus22-pert}.
Together with the boundary conditions \eqref{bound_cond_pert_compl},
this system defines an eigenvalue problem for the frequencies $\chi^2$.
The question of stability is thus reduced to a study of the possible
values of $\chi^2$.
Whenever any values of $\chi^2$ are found to be negative,
then the corresponding perturbations will grow,
and the configurations in question will be unstable
against radial oscillations.

In performing the stability analysis,
we also require that the radial perturbations do not change
the total particle number $N$, given by the expression \eqref{tot_part_number}
\cite{Jetzer:1991jr}.
This means that the perturbation of the total particle number
$$ N_p =8\pi \omega \int_0^r e^{\mu_0-\nu_0/2} \phi_0^2\left[
2\mu_p+\frac{1}{\phi_0}\left(2\Phi_{1_p}-\frac{1}{\omega}\dot{\Phi}_{2p}\right)
\right]dr^\prime
$$
should be equal to zero.
Substituting the perturbed solutions obtained from
Eqs.~\eqref{sf-real-pert}, \eqref{00plus11-pert}, and \eqref{00plus22-pert}
into this expression,
we have to check that the resulting perturbations
indeed satisfy the condition $N_p=0$.

The results of the calculations of the lowest eigenvalue $\chi_0^2$
are shown in Fig.~\ref{fig_chi}.
Starting  with the case of a real scalar field,
i.e., for  boson frequency $\Omega=0$ and $\Phi_2=0$, we see in the left panel
that in this limiting case the lowest eigenvalue $\chi_0^2$
remains negative for all values of the self-coupling constant $\Lambda$.
Thus, all corresponding systems are unstable.
For the case of a time-dependent scalar field with a finite boson frequency
$\Omega\neq 0$, the instability becomes even worse.
For any given $\Lambda$, as $\Omega$ increases,
the eigenvalue $\chi_0^2$ decreases,
as demonstrated in the right panel of Fig.~\ref{fig_chi}, where,
using two values of $\Lambda$ as examples,
the typical dependence of $\chi_0^2$ on the central value
of the scalar field $\varphi_c$ is shown.
Thus, the considered topologically nontrivial configurations
are always unstable against linear radial perturbations.

\section{Conclusion}
\label{concl}

We have considered equilibrium configurations
with a nontrivial wormholelike spacetime topology
supported by the complex ghost scalar field
with a quartic potential of the Mexican hat form \eqref{poten_Mex_complex}.
This potential has allowed us to find regular asymptotically flat solutions
for an explicitly time-dependent complex scalar field,
oscillating with a frequency $\Omega$.
We have shown that, depending on the values of $\Omega$
and the self-coupling constant $\Lambda$,
these solutions describe configurations with finite sizes,
which may possess a positive or a negative  ADM mass,
as seen in Fig.~\ref{fig_M_sigma}.

However, the mode stability analysis against linear perturbations
has revealed an instability of these systems
with respect to radial perturbations possessing reflection symmetry.
(Note that in order
to demonstrate the instability of the system
 it is sufficient to
consider only this particular subset of perturbations,
although one can consider  also other types of perturbations,
including those given, for example, only on one side of the throat.)
 We recall that the
instability is already present in the limiting case of
configurations supported by a real scalar field
(with boson frequency $\Omega=0$),
and  here we have shown that it
is aggravated in the case of an oscillating complex scalar field.
In particular, for the complex field, the instability increases
with increasing boson frequency $\Omega$,
where for any given $\Lambda$ the square of the lowest eigenfrequency
of the radial oscillations becomes increasingly negative
(as seen in Fig.~\ref{fig_chi}).
Thus, as in the case of real ghost scalar fields
considered in the literature before
\cite{Shinkai:2002gv,Gonzalez:2008wd,Gonzalez:2008xk,Bronnikov:2011if,Bronnikov:2012ch,Sarbach:2012wi},
the use of a complex ghost scalar field does not allow for
stable solutions, as demonstrated here for
a Mexican hat type potential \eqref{poten_Mex_complex}.

\section*{Acknowledgments}
V.D. and V.F. gratefully acknowledge support provided by Grant No.~$\Phi.0755$
in Fundamental Research in Natural Sciences
by the Ministry of Education and Science of the Republic of Kazakhstan.
V.F. acknowledges financial support within the
DAAD program Research Stays for University
Academics and Scientists and also would like to thank the
Carl von Ossietzky University of Oldenburg for hospitality
while this work was carried out.
B.K. and J.K. gratefully acknowledge support
by the DFG Research Training Group 1620 {\sl Models of Gravity}
as well as by FP7, Marie Curie Actions, People,
International Research Staff Exchange Scheme (Grant No.~IRSES-606096).

\end{document}